# Strategic Environmental Corporate Social Responsibility (ECSR) Certification and Endogenous Market Structure

Ajay Sharma

Indian Institute of Management, Indore (India)

Siddhartha K. Rastogi

Indian Institute of Management, Indore (India)

**Correspondence address:**

**Ajay Sharma**, J-206, Academic Block, Indian Institute of Management Indore, Prabandh Shikhar, Rau-Pithampur Road, Indore, M.P. (India) - 453556. Ph: +91-7312439622. E-mail: ajays@iimidr.ac.in; ajaysharma87@gmail.com.

**Siddhartha K. Rastogi**, B-101, Academic Block, Indian Institute of Management Indore, Rau-Pithampur Road, Indore, M.P. (India) - 453556. Ph: +91-7312439534. E-mail: srastogi@iimidr.ac.in

# Strategic Environmental Corporate Social Responsibility (ECSR) Certification and Endogenous Market Structure

Abstract

This paper extends the findings of *Liu et al. (2015, Strategic environmental corporate social responsibility in a differentiated duopoly market, Economics Letters)*, along two dimensions. First, we consider the case of endogenous market structure *a la* Vives and Singh (1984, *Price and quantity competition in a differentiated duopoly, The Rand Journal of Economics*). Second, we refine the ECSR certification standards in differentiated duopoly with rankings. We find that optimal ECSR certification standards by NGO are the highest in Bertrand competition, followed by mixed markets and the lowest in Cournot competition. Next, NGO certifier will set the ECSR standards below the optimal level. Also, we show that given the ECSR certification standards, there is a possibility of both price and quantity contracts choices by the firms in endogenous market structure.

1. Introduction

Corporate Social Responsibility (CSR) has become a mainstream pursuit among the business activities of firms in the past few years, wherein more than 30% (71% and 90%) of companies in the US (the UK and Japan, respectively) adopted CSR reporting in 2013 (Kim *et al.,* 2017*)*.

Given the strategic importance of CSR activities as a non-core business pursuit and their significant implication for costs, eco-labeling, certification, hallmarking etc. are the common ways of CSR signaling especially for environmental outcomes. Though certification is not a perfect mechanism, it is sufficiently trustworthy to convey useful information (Auriol and Schilizzi, 2015).

The certification can come from self or third-party and can be mandatory or optional. The existing literature on the strategic aspects of third-party certification focuses on nature of competition and third-party certifiers. Manasakis *et al.* (2013) suggest that the certification by alternative third parties differ with respect to their objectives and has implications for certification standards. Liu *et al.* (2015) compares the ECSR certification level in Cournot versus Bertrand competition and show that certification standards are lower in Bertrand than Cournot competition.

Our contribution to this literature is two folds. First, we extend the analysis of Liu *et al.* (2015) by endogenizing the market structure *a la* Singh and Vives (1984). If the firms have option of price or quantity contracts, given the ECSR standards, then, what would be optimal choice for the firms? Second, we refine the ECSR certification standards in this endogenous market structure by providing rankings and then considering uniform standards.

2. The Model

Based on Manasakis *et al.* (2013) and Liu *et al.* (2015), the utility function of a representative consumer is

$$U = (A + e_1 \alpha s_1)q_1 + (A + e_2 \alpha s_2)q_2 - \frac{(q_1^2 + 2\gamma q_1 q_2 + q_2^2)}{2}$$

where $q_i$ is output and $s_i$ is the level of ECSR, for firm $i$ ($i = 1,2$). The parameter $\gamma \in (0,1)$ measures the nature of products being substitutes ($\gamma > 0$). The parameter $\alpha \in (0,1)$ indicates the consumer's preference for firm's ECSR activities. The firms choose ECSR as a strategic variable. Based on Manasakis *et al.* (2013), we consider that ECSR activities can be informed to consumers through a credible signal. For the same, the firm seeks certification from a third-party NGO certifier who maximizes Net Consumer Surplus (NCS). A firm can get the certification $e_i$ if it satisfies criteria of minimum level of ECSR activities $\underline{s}$ :

$$e_i = \begin{cases} 0 \text{ if } s_i < \underline{s} \text{ and firm } i \text{ does not receive a ECSR cetification} \\ 1 \text{ if } s_i \geq \underline{s} \text{ and firm } i \text{ receives a ECSR cetification} \end{cases}$$

It is important to note that a firm will consider doing ECSR activity only if it generates net positive benefits. A firm would spend at $\underline{s}$ (minimum ECSR for certification) and not beyond that. If both firms choose to get the certification by spending $\underline{s}$ on ECSR, the representative consumer's utility would be:

$$U = (A + e_1 \alpha \underline{s})q_1 + (A + e_2 \alpha \underline{s})q_2 - \frac{(q_1^2 + 2\gamma q_1 q_2 + q_2^2)}{2}$$

The corresponding demand functions would be $q_i = \frac{A(1-\gamma) - p_i + \gamma p_j + \alpha e_i \underline{s} - \alpha \gamma e_j \underline{s}}{1-\gamma^2}$ and inverse demand functions $p_i = A - q_i - \gamma q_j + \alpha e_i \underline{s}$, for $i, j = 1,2; i \neq j$.

We assume that firms use same technology with cost of production as zero, without loss of generality. Also, one unit of output produces one unit of pollution emission. The NGO certifier will not charge any fee for certification if firm complies with ECSR standards. The cost of ECSR for firms is $s_i^2$.

The firm's profit function is, $\pi_i = p_i q_i - e_i \underline{s}^2$, $i = 1,2$. NGO certifier's objective function is $NCS = CS - \frac{d(q_1 + q_2 - e_1 \underline{s} - e_2 \underline{s})^2}{2}$ where $CS = \frac{(q_1^2 + 2\gamma q_1 q_2 + q_2^2)}{2}$; and $d > 0$ is the marginal environmental damage due to emissions.

3. The Game

The game is organized as follows. In the *first stage*, the firm decides to choose price or quantity contracts. In the *second stage*, the certifier decides threshold level of ECSR for certification. Firms meeting the threshold condition get the certification, otherwise not. In the *third stage*, firms choose the level of output and prices to maximize their profits.

We solve the game using backward induction.

*3.1. Product market competition*

In this stage, we analyze the four possible options: a) both firms choose prices (*pp*) *i.e.,* Bertrand competition; b) both firms choose quantities (*qq*) *i.e.,* Cournot competition; c) one firm chooses price (quantity) contract while the other firm chooses the quantity (price) contract *i.e.*, *pq* (*qp*) outcomes.

We avoid providing the calculations for (a) and (b) option for the sake of brevity, as they are identical to Liu et al. (2015). Please refer to the online appendix for the same.

***Proposition 1:*** *The NGO certifier will set the standards,* $\underline{s} = \underline{s}^{PPU}$ *and* $\underline{s}^{QQU}$ *in the Bertrand (pp game) and Cournot (qq game) respectively.*

Proof: See online appendix.

Next, both (c) and (d) will be identical in nature. Therefore, we only solve the *pq game.*

*pq game (Price versus Quantity Contract)*

We use the superscript PQ for price-quantity contract case *i.e.*, firm 1 decides price while firm 2 decides quantity. The outcomes in the product market with firms not adopting ECSR are

$$q_1^{PQN} = \frac{A(2-\gamma-\gamma^2)}{4-3\gamma^2}; \; q_2^{PQN} = \frac{A(2-\gamma)}{4-3\gamma^2}; p_1^{PQN} = \frac{A(2-\gamma-\gamma^2)}{4-3\gamma^2}; \; p_2^{PQN} = \frac{A(2-\gamma)(1-\gamma)(1+\gamma)}{4-3\gamma^2}; \; \pi_1^{PQN} = \frac{A^2(2-\gamma-\gamma^2)^2}{(4-3\gamma^2)^2}; \pi_2^{PQN} = \frac{A^2(2-\gamma)^2(1-\gamma^2)}{(4-3\gamma^2)^2}; \; NCS^{PQN} = \frac{A^2(8-10\gamma^2+3\gamma^4-d(4-\gamma(2+\gamma))^2)}{2(4-3\gamma^2)^2} \quad (5)$$

If the firms choose to opt for ECSR activities and get certification, the equilibrium outcomes would be,

$$q_1^{PQC} = \frac{(2-\gamma-\gamma^2)(A+\alpha\underline{s})}{4-3\gamma^2}; \; q_2^{PQC} = \frac{(2-\gamma)(A+\alpha\underline{s})}{4-3\gamma^2}; p_1^{PQC} = \frac{(2-\gamma-\gamma^2)(A+\alpha\underline{s})}{4-3\gamma^2}; \; p_2^{PQC} = \frac{(2-\gamma)(1-\gamma)(1+\gamma)(A+\alpha\underline{s})}{4-3\gamma^2}; \; \pi_1^{PQC} = \frac{(A^2+2A\alpha\underline{s}+\alpha^2\underline{s}^2)(2-\gamma-\gamma^2)^2-(4-3\gamma^2)^2\underline{s}^2}{(4-3\gamma^2)^2}; \pi_2^{PQC} \frac{(A^2+2A\alpha\underline{s}+\alpha^2\underline{s}^2)(2-\gamma)^2(1-\gamma^2)+((4-3\gamma^2)^2)}{(4-3\gamma^2)^2}; \; NCS^{PQC} = \frac{(2-\gamma^2)(A+\alpha\underline{s})^2}{8-6\gamma^2} - \frac{A^2d(-4+\gamma(2+\gamma))^2}{2(4-3\gamma^2)^2}$$

For $q_1^{PQC} > \underline{s}$, $\underline{s} < \frac{2A-A\gamma-A\gamma^2}{4-2\alpha+\alpha\gamma-3\gamma^2+\alpha\gamma^2}$ and for $q_2^{PQC} > \underline{s}$, $\underline{s} < \frac{2A-A\gamma}{4-2\alpha+\alpha\gamma-3\gamma^2}$ should be satisfied. For both $q_1^{PQC} > \underline{s}$ and $q_2^{PQC} > \underline{s}$, $\underline{s} < \frac{2A-A\gamma-A\gamma^2}{4-2\alpha+\alpha\gamma-3\gamma^2+\alpha\gamma^2}$ must be satisfied. Further $q_2^{PQC} > q_1^{PQC}$ holds for all parametric values.

Firm 1 would be willing to adopt ECSR certification if $\pi_1^{PQC} > \pi_1^{PQN}$ i.e., $\underline{s} < \underline{s}^{PQU1} = \frac{2A-A\gamma-A\gamma^2}{4-2\alpha+\alpha\gamma-3\gamma^2+\alpha\gamma^2}$ holds. For firm 2, decision to adopt ECSR certification is chosen if $\pi_2^{PQC} > \pi_2^{PQN}$ i.e. $\underline{s} < \underline{s}^{PQU2} = \frac{A\alpha(2-\gamma-\gamma^2)^2}{(4-3\gamma^2)^2-\alpha^2(2-\gamma-\gamma^2)^2}$ holds.

Also, comparing the upper threshold of spending on ECSR, we observe that firm 1 (choosing price) has higher threshold than firm 2 (choosing quantity)'s ECSR spending, *i.e.*, $\underline{s}^{PQU1} > \underline{s}^{PQU2}$.

***Lemma 1:*** *In a price vs. quantity game, price setting firm has higher threshold for ECSR spending than quantity setting firm.*

*NGO certifier*

Coming to second stage, we obtain the optimal choice of ECSR certification standard for NGO certifier by evaluating $\frac{d\ NCS^{PQC}}{ds} = 0$. We get

$$\underline{s}^{PQ*} = \frac{A(\alpha(8-10\gamma^2+3\gamma^4)-d(-4+\gamma(2+\gamma))(8-6\gamma^2+\alpha(-4+\gamma(2+\gamma))))}{\alpha^2(-8+10\gamma^2-3\gamma^4)+d(8-6\gamma^2+\alpha(-4+\gamma(2+\gamma)))^2}$$

$\underline{s}^{PQ*} > 0$ if $d > \frac{\alpha^2(8-10\gamma^2+3\gamma^4)}{(8-6\gamma^2+\alpha(-4+\gamma(2+\gamma)))^2}$ holds.

Further, $\underline{s}^{PQ*} > \underline{s}^{PQU1}$ and $\underline{s}^{PQ*} > \underline{s}^{PQU2}$ when $d > \frac{\alpha^2(8-10\gamma^2+3\gamma^4)}{(8-6\gamma^2+\alpha(-4+\gamma(2+\gamma)))^2}$. This means that certifier's optimal level of ECSR standard would be higher than the upper limit for the firms in price *vs.* quantity competition and any firm will not spend on ECSR if a certifier sets the standard at $\underline{s}^{PQ*}$. NGO certifier can set the ECSR standard for certification at either $\underline{s} = \underline{s}^{PQU1}$ or $\underline{s} = \underline{s}^{PQU2}$ level for participation. If $\underline{s} = \underline{s}^{PQU1}$ is chosen as ECSR standard, then only price-setting firm 1 will get the certification and quantity-setting firm 2 will not get ECSR certification. Interestingly, profit of firm 2 will be higher than firm 1.

On the other hand, $\underline{s} = \underline{s}^{PQU2}$ as ECSR standard leads to both firms getting the certification. In this case also, firm 1's profit would be lower than firm 2's.

This indicates that quantity setting firm 2 has net advantage over price setting firm 1 irrespective of whether firm 1 unilaterally get the ECSR certification or both firms get the certification. This is a new result.

Therefore, to induce the firms in adopting the certification, the standard would be set at $\underline{s} = \underline{s}^{PQU}$. Further, we find that consumers and firms would benefit from such ECSR standard as compared to no ECSR at all because $NCS^{PQC} > NCS^{PQN}$.

***Proposition 2:*** *In a price vs. quantity competition,*

   a) *NGO certifier will set the ECSR standard below the optimal level*
   b) *if ECSR certification standard is set at $\underline{s} = \underline{s}^{PQU1}$, then only price-setting firm will get the certification, whereas quantity-setting firm 2 will not opt for certification.*
   c) *if ECSR certification standard is set at $\underline{s} = \underline{s}^{PQU2}$, then both firms will get the certification and it is beneficial for both firms and consumers.*

4. Comparison of ECSR Certification Standards

Comparing the optimal ECSR standard, $\underline{s}$ by the NGO certifier across endogenous market structure, we observe that $\underline{s}^{PP*} > \underline{s}^{PQ*} > \underline{s}^{QQ*}$ indicating Bertrand has the highest level followed by price-quantity and lastly Cournot case.

***Proposition 3:*** *Across the spectrum of market structure, the NGO certifier's optimal ECSR standard rankings are $\underline{s}^{PP*} > \underline{s}^{PQ*} > \underline{s}^{QQ*}$.*

In all the cases, the NGO certifier is not able to implement the optimal level of ECSR standard because the firms will not adopt such ECSR standards as that leads to lower profit for them. Therefore, the certifier would choose a sub-optimal ECSR standard to incentivize the firms. Comparing these equilibrium standards, we get the ranking in proposition 4.

***Proposition 4:*** *The NGO certifier's equilibrium ECSR standard rankings are $\underline{s}^{PQU1} > \underline{s}^{QQU} > \underline{s}^{PPU} > \underline{s}^{PQU2}$.*

5. Endogenous Market Structure: Price or Quantity Contract

Now, we solve the first stage of the game where firms have options to choose price or quantity contracts in the product market competition. For the sake of brevity, we do not consider the case where no firm chooses ECSR certification. The outcome of that subgame will be identical to Singh and Vives (1984).

***Lemma 2 (*Singh and Vives, 1984*):*** *In a product market competition for substitute goods[1], with price and quantity as strategic choices, firms choose quantity contracts as dominant strategies.*

*ECSR certification standards and market structure*

If firms opt for the ECSR certification, then the outcome of the subgame can differ from Singh and Vives (1984). The certifier can choose a uniform standard irrespective of the nature of market competition, or different standards based on nature of competition[2]. We only consider possibility of uniform ECSR certification standards.

In case of uniform ECSR standard, there are four choices, $\underline{s}^{PQU1} > \underline{s}^{QQU} > \underline{s}^{PPU} > \underline{s}^{PQU2}$ (see Proposition 4). If the NGO certifier sets the lowest three ECSR certification standard, *i.e.*, either $\underline{s}^{PQU2}$ or $\underline{s}^{PPU}$ or $\underline{s}^{QQU}$, then the Nash equilibrium outcome of the game in Table 1, is {Quantity, Quantity}. On the other hand, if the ECSR certifier sets the standard at the highest level possible *i.e.*, $\underline{s}^{PQU1}$, then there are two Nash equilibria outcomes of the game {Price, Quantity} and {Quantity, Price}.

---

[1] In this paper, we only consider substitute goods in the market offered by competing firms.
[2] But from a real-world point of view, such standards may not be feasible due to monitoring issues, discrimination, and mimicking behavior among firms.

| Table 1: Price-Quantity Contract Game (with ECSR certification) | | | |
|---|---|---|---|
| | | Firm 2 | |
| | | Price | Quantity |
| Firm 1 | Price | $\pi_1^{PPC}, \pi_2^{PPC}$ | $\pi_1^{PQC}, \pi_2^{PQC}$ |
| | Quantity | $\pi_1^{QPC}, \pi_2^{QPC}$ | $\pi_1^{QQC}, \pi_2^{QQC}$ |

**Proposition 5:** *In a price-quantity contract game,*

a) *If a certifier decides, the uniform ECSR standard at either $\underline{s}^{PQU2}$ or $\underline{s}^{PPU}$ or $\underline{s}^{QQU}$ level, the subgame perfect Nash equilibrium is {Quantity, Quantity}*
b) *If the certifier decides, the uniform ECSR standard at $\underline{s}^{PQU1}$, there are two subgame perfect Nash equilibria {price, Quantity}, {Quantity, Price}*

Proof: See online appendix

6. Conclusion

In this paper, we analyze the relationship between endogenous market structure and strategic ECSR in a differentiated duopoly. We show that NGO certifier will always set the ECSR standards below the optimal level to ensure participation. In a price-quantity game, there is possibility of partial or full compliance with ECSR standards. Lastly, while setting a uniform ECSR standards in endogenous market structure, there is a possibility of Cournot outcome as well as mixed market outcome.

A1. A $pp$ game (Bertrand Competition)

We use the superscript PPN to denote equilibrium outcome for firms not adopting ECSR in $pp$ game *i.e.*, Bertrand competition, otherwise PPC. Solving the game, we get

$$q_1^{PPN} = q_2^{PPN} = \frac{A}{2+\gamma-\gamma^2} \; ; p_1^{PPN} = p_2^{PPN} = \frac{A(1-\gamma)}{2-\gamma}; \; \pi_1^{PPN} = \pi_2^{PPN} = \frac{A^2(1-\gamma)}{(2-\gamma)^2(1+\gamma)} \; ; \text{NCS}^{PPN} = \frac{A^2(1-2d+\gamma)}{(2+\gamma-\gamma^2)^2} \quad (1)$$

If the firms decide to adopt for ECSR and get certification for the same, the outcomes are

$$q_1^{PPC} = q_1^{PPC} = \frac{A + \alpha \underline{s}}{2 + \gamma - \gamma^2} \; ; p_1^{PPC} = p_2^{PPC} = \frac{(1-\gamma)(A + \alpha \underline{s})}{2-\gamma}; \; \pi_1^{PPC} = \pi_2^{PPC}$$

$$= \frac{(1-\gamma)(A + \alpha \underline{s})^2}{(2-\gamma)(2+\gamma-\gamma^2)} - \underline{s}^2 \; ;$$

$$\text{NCS}^{PPC} = \frac{(A+\alpha\underline{s})^2}{(2-\gamma)^2(1+\gamma)} - \frac{2d(A+(\alpha+(-2+\gamma)(1+\gamma))\underline{s})^2}{(2+\gamma-\gamma^2)^2} \quad (2)$$

For a firm to be adopting ECSR, the certification threshold needs to be lower than the level of pollution, otherwise the cost will be more than its benefits. For $q_i^{PPC} > \underline{s}, \underline{s} < \frac{A}{2-\alpha+\gamma-\gamma^2}$ should be satisfied.

Comparing the profits of firms[3] with or without ECSR,

$$\pi_1^{PPC} - \pi_1^{PPN} = \frac{A^2(1-\gamma) + 2A\alpha(1-\gamma)\underline{s} - (4 - \alpha^2(1-\gamma) - (3-\gamma)\gamma^2)\underline{s}^2}{(2-\gamma)^2(1+\gamma)}$$

---
[3] Given the symmetry of the firms and their outcomes, we only compare the results of one firm, and it holds for both of them.

We observe that $\pi_1^{PPC} - \pi_1^{PPN} > 0$ if $\underline{s} < \underline{s}^{PPU} = \frac{2A\alpha(1-\gamma)}{4-\alpha^2+\alpha^2\gamma-3\gamma^2+\gamma^3}$, which provides the upper bound for the ECSR spending to adopt the ECSR certification. So, firms will spend strategically on ECSR and get certification if $\underline{s} < \underline{s}^{PPU}$.[4]

*Optimal ECSR Certification Standard*

We obtain the optimal choice of ECSR certification standard in case of NGO certifier by evaluating $\frac{d\ NCS^{PPC}}{ds} = 0$. We get,

$$\underline{s}^{PP*} = \frac{A(\alpha + \alpha\gamma - 2d(\alpha - (2-\gamma)(1+\gamma)))}{2d(\alpha - (2-\gamma)(1+\gamma))^2 - \alpha^2(1+\gamma)}$$

$\underline{s}^{PP*} > 0$ if $d > \frac{\alpha^2+\alpha^2\gamma}{2(\alpha-(2-\gamma)(1+\gamma))^2}$ holds. Further, $\underline{s}^{PP*} > \underline{s}^{PPU}$ when $d > \frac{\alpha^2+\alpha^2\gamma}{2(\alpha-(2-\gamma)(1+\gamma))^2}$. This means that an NGO certifier's optimal level of ECSR standard would be higher that the upper limit for the firms in Bertrand competition and a firm would not choose to spend on ECSR if a certifier sets the standard at $\underline{s}^{PP*}$. Therefore, to induce the firms, the standard would be set at $\underline{s} = \underline{s}^{PPU}$ by the NGO certifier. Further, we can also show that consumer and firms would benefit from such ECSR standard as compared to no ECSR at all *i.e.,* $NCS^{PPC} > NCS^{PPN}$.

***Proposition A1:*** *An NGO certifier would set the ECSR standard below the optimal level if firms engage in Bertrand competition and, it is beneficial for both firms and consumers in terms of profit and net consumer surplus, respectively.*

A2. A $qq$ game (Cournot Competition)

---

[4] Superscript U denotes upper bound.

For Cournot game, we use the superscript QQ. The outcomes of the product market competition, if firms do not adopt ECSR are, $q_1^{QQN} = q_2^{QQN} = \frac{A}{2+\gamma}$; $p_1^{QQN} = p_2^{QQN} = \frac{A}{2+\gamma}$; $\pi_1^{QQN} = \pi_2^{QQN} = \frac{A^2}{(2+\gamma)^2}$; $NCS^{QQN} = \frac{A^2(1-2d+\gamma)}{(2+\gamma)^2}$ (3)

On the other hand, if the firms decide to adopt ECSR certification, the outcomes would be,

$q_1^{QQC} = q_2^{PPC} = \frac{A+\alpha\underline{s}}{2+\gamma}$; $p_1^{QQC} = p_2^{QQC} = \frac{A+\alpha\underline{s}}{2+\gamma}$; $\pi_1^{QQC} = \pi_2^{QQC} = \frac{(A+\alpha\underline{s})^2}{(2+\gamma)^2} - \underline{s}^2$; $NCS^{QQC} = \frac{(1+\gamma)(A+\alpha\underline{s})^2 - 2d(A-(2-\alpha+\gamma)\underline{s})^2}{(2+\gamma)^2}$ (4)

For $q_i^{QQC} > \underline{s}$, $\underline{s} < \frac{A}{2-\alpha+\gamma}$ should be satisfied. Further comparing the profits of firms with and without adopting ECSR,

$$\pi_1^{QQC} - \pi_1^{QQN} = \frac{A^2 + 2A\alpha\underline{s} + (-2+\alpha-\gamma)(2+\alpha+\gamma)\underline{s}^2}{(2+\gamma)^2}$$

A firm would profit from adopting ECSR if $\pi_1^{QQC} > \pi_1^{QQN}$ i.e., when $\underline{s} < \underline{s}^{QQU} = \frac{2A\alpha}{4-\alpha^2+4\gamma+\gamma^2}$. This denotes the upper bound to spend on ECSR for certification.

*Optimal ECSR Certification Standard*

We obtain the optimal choice of ECSR certification standard in case of NGO certifier by evaluating $\frac{d\,NCS^{QQC}}{ds} = 0$. We get,

$$\underline{s}^{QQ*} = \frac{A(\alpha + \alpha\gamma + 2d(2-\alpha+\gamma))}{2d(2-\alpha+\gamma)^2 - \alpha^2(1+\gamma)}$$

$\underline{s}^{QQ*} > 0$ if $d > \frac{\alpha^2+\alpha^2\gamma}{2(2-\alpha+\gamma)^2}$ holds. Further, $\underline{s}^{QQ*} > \underline{s}^{QQU}$ when $d > \frac{\alpha^2+\alpha^2\gamma}{2(2-\alpha+\gamma)^2}$. This means that a certifier's optimal level of ECSR standard would be higher than the upper limit for the firms in Cournot competition and a firm would not choose to spend on ECSR if a certifier sets the standard at $\underline{s}^{QQ*}$. Therefore, to induce the firms, the standard would be set at $\underline{s} = \underline{s}^{QQU}$ by the

NGO certifier. Further, we can also show that consumer and firms would benefit from such ECSR standard as compared to no ECSR at all *i.e.*, $\text{NCS}^{QQC} > \text{NCS}^{QQN}$.

***Proposition A2:*** *An NGO certifier would set the ECSR standard below the optimal level if firms engage in Cournot competition and, it is beneficial for both firms and consumers in terms of profit and net consumer surplus, respectively.*

### A3. Proof for Proposition 5

Proof:

a) In all three cases, we observe that $\pi_1^{QPC} > \pi_1^{PPC}$ and $\pi_1^{QQC} > \pi_1^{PQC}$ for firm 1; and $\pi_2^{PQC} > \pi_2^{PPC}$ and $\pi_2^{QQC} > \pi_2^{QPC}$ for firm 2. This makes 'Quantity contract' as the dominant strategy for both the firms and the Nash equilibrium is {Quantity, Quantity}.

b) In this case, we observe that $\pi_1^{QPC} > \pi_1^{PPC}$ and $\pi_1^{QQC} < \pi_1^{PQC}$ for firm 1; and $\pi_2^{PQC} > \pi_2^{PPC}$ and $\pi_2^{QQC} < \pi_2^{QPC}$ for firm 2. This means that there are no dominant strategies and the best response of either firm is the choice of opposite strategy, and the Nash equilibria are {Price, Quantity}, {Quantity Price}.